\documentclass{article}[12pt]
\usepackage{amssymb}
\usepackage{amsmath}

\newtheorem{theorem}{Theorem}[section]

\newtheorem{definition}{Definition}[section]
\numberwithin{equation}{section}

\begin{document}

\centerline{Accardi complementarity for $-1/2 < \mu <0$
and related results}

\vskip .8cm

\centerline{Lenin Augusto Echavarr\'ia Cepeda\footnote{Research partially supported
by CONACYT (Mexico) project 49187 and CONACYT (Mexico) student grant 165360 }}
\centerline{Centro de Investigaci\'on en Matem\'aticas, A.C.(CIMAT)}
\centerline{Guanajuato, Mexico}
\centerline{email: lenin@cimat.mx}

\vskip .5cm

\centerline{Claudio de Jes\'{u}s Pita Ruiz Velasco$^2$}
\centerline{Universidad Panamericana}
\centerline{Mexico City, Mexico}
\centerline{email: cpita@mx.up.mx}

\vskip .5cm

\centerline{Stephen Bruce Sontz\footnote{Research partially
supported by CONACYT (Mexico) project 49187.}}
\centerline{Centro de Investigaci\'on en Matem\'aticas, A.C.(CIMAT)}
\centerline{Guanajuato, Mexico}
\centerline{email: sontz@cimat.mx}

\vskip .5cm

\begin{abstract}
\noindent
We show that the momentum and position operators of
$\mu$-deformed quantum mechanics for $-1/2 < \mu < 0$ are not Accardi complementary.
This proves an earlier conjecture of the last two authors as well as extending
their analogous result for the case $\mu >0$.
We also prove some related formulas that were conjectured by the same authors.
\end{abstract}

\section{Introduction}

In this article we present a new result in the same direction as the main result
of the recent work \cite{PS3} as well as proving some formulas that were
also conjectured there.
This article should be considered as a sequel to \cite{PS3}.
For the reader's convenience, we collect in this
section some of the basic material in \cite{PS3}.

First we present some relevant facts of the so-called
{\em $\mu$-deformed quantum mechanics}.
For more details, refer to \cite{CAASBS1}, \cite{CAASBS2}, \cite{MA},
\cite{PS}, \cite{PS2}, \cite{PS3} and \cite{RO}.
In this theory the
mathematical objects of quantum mechanics (position and momentum operators,
configuration space, phase space, etc.) are deformed by a parameter
$\mu >-\frac{1}{2}$ (the undeformed theory corresponding to $\mu =0$).
We will be dealing with the complex Hilbert space
$L^{2}\left( {\mathbb{R}},m_{\mu }\right) $,
where the measure $m_{\mu }$ is given by
$$
   dm_{\mu}(x):=\left( 2^{\mu +\frac{1}{2}}\Gamma \left( \mu +\frac{1}{2}\right)
   \right) ^{-1}|x|^{2\mu }dx
$$
for $x\in {\mathbb{R}}$.
Here $dx$ is Lebesgue measure on ${\mathbb{R}}$
and $\Gamma $ is the Euler gamma function.
The normalization of this measure is chosen to give us a self-dual
($\mu$-deformed) Fourier transform. See \cite{RO} for details.
In this Hilbert space $L^{2}\left( {\mathbb{R}},m_{\mu }\right) $
we have two
unbounded self-adjoint operators: the $\mu $-deformed position
operator $Q_{\mu }$
and the $\mu $-deformed momentum operator $P_{\mu }$.
These are defined for
$x\in {\mathbb{R}}$ and certain elements
$\psi \in L^{2}({\mathbb{R}},m_{\mu })$ by
\begin{eqnarray*}
Q_{\mu }\psi (x):= &&x\psi (x), \\
P_{\mu }\psi (x):= &&\frac{1}{i}\left( \psi ^{\prime }(x)+\frac{\mu }{x}
(\psi (x)-\psi (-x)\right) .
\end{eqnarray*}
We omit details about exact domains of definition.
Interest in these
operators originates in Wigner \cite{wig} where equivalent forms of them are used
as examples of operators that do not satisfy the usual canonical commutation
relation in spite of the fact that they do satisfy the equations of motion
$i[H_{\mu },Q_{\mu }]=P_{\mu }$ and $i[H_{\mu },P_{\mu }]=-Q_{\mu }$ for the
Hamiltonian $H_{\mu }:=\frac{1}{2}(Q_{\mu }^{2}+P_{\mu }^{2})$.
What does
hold is the $\mu $-deformed canonical commutation relation:
$i[P_{\mu},Q_{\mu }]=I+2\mu J$, where $I$ is the identity operator and $J$ is the
parity operator $J\psi (x):=\psi (-x)$.

In \cite{acc} Accardi introduced a definition of complementary observables in
quantum mechanics.
We now generalize that definition to the current context.
We use the usual identification of observables in quantum mechanics as
self-adjoint operators acting in some Hilbert space.

\bigskip

\begin{definition}

We say that the (not necessarily bounded) self-adjoint operators $S$
and $T$ acting in $L^{2}(\mathbb{R},m_{\mu })$
are {\em Accardi complementary} if for any pair of bounded Borel subsets $A$
and $B$ of $\mathbb{R}$ we have that the
operator $E^{S}(A)E^{T}(B)$ is trace class with trace given by
\begin{equation*}
Tr\left( E^{S}(A)E^{T}(B)\right) =m_{\mu }(A)m_{\mu }(B).
\end{equation*}

\end{definition}

\bigskip

Here $E^S$ is the projection-valued measure on ${\mathbb{R}} $ associated
with the self-adjoint operator $S$ by the spectral theorem, and similarly
for $E^T$.

So, $E^{S}(A)E^{T}(B)$ is clearly a bounded operator acting on
$L^{2}({\mathbb{R}},m_{\mu })$.
But whether it is also trace class is another
matter.
And, given that it is trace class, it is a further matter to
determine if the trace can be written as the product of measures, as
indicated.
Accardi's result in \cite{acc} (which is also discussed in detail and
proved in \cite{cas-var}) is that $Q\equiv Q_{0}$ and $P\equiv P_{0}$ are Accardi
complementary.
Accardi also conjectured that this property of $Q$ and $P$
characterized this pair of operators acting on $L^{2}({\mathbb{R}},m_{0})$.
It turns out that this is not so.
(See \cite{cas-var}.)

The main result in \cite{PS3} is the following theorem.

\bigskip
\begin{theorem}
\label{thm1}
Let $A$ and $B$ be bounded
Borel subsets of $\mathbb{R}$ with $0\notin A^{-}$, the closure of $A$.
Then
$E^{Q_{\mu }}(A)E^{P_{\mu }}(B)$ is a trace class operator in
$L^{2}(\mathbb{R},m_{\mu })$ for any $\mu >-\frac{1}{2}$ with
\begin{equation}
\label{finite_trace}
0\leq Tr\left( E^{Q_{\mu }}(A)E^{P_{\mu }}(B)\right) =\int_{A}dm_{\mu
}(x)\int_{B} dm_{\mu }\left( k\right) |\exp_{\mu }(ikx)|^{2} <\infty .
\end{equation}

Moreover, if $\mu >0$ and $m_{\mu }(A) m_{\mu }(B)\neq 0$ then we have that
\begin{equation}
Tr\left( E^{Q_{\mu }}(A)E^{P_{\mu }}(B)\right) < m_{\mu }(A) m_{\mu }(B).
\end{equation}

In particular, the operators $Q_{\mu }$ and $P_{\mu }$
are not Accardi complementary if $\mu >0$.
\end{theorem}

In \cite{PS3} the conjecture is made that
\begin{gather}
\label{conj}
Tr \left( E^{Q_\mu}(A) E^{P_\mu}(B) \right)
> m_\mu(A) m_\mu(B)
\end{gather}
for $A,B$ bounded Borel sets of positive $m_\mu$ measure and $-1/2 < \mu < 0$.
We shall prove this result under the extra technical hypothesis $0\notin A^{-}$
and thereby establish that the operators $Q_{\mu }$ and $P_{\mu }$
are not Accardi complementary for $-1/2 < \mu <0$.

The organization of this article is a follows.
In the next section we present a theorem that
will allow us to prove the main result in Section~3.
Finally in Section~4, we prove several new identities
of $\mu$-deformed quantities, including all
of those conjectured in \cite{PS3}.

\bigskip

\section{Preliminary Results}

\bigskip

We take $\mu >-\frac{1}{2}$ arbitrary unless otherwise stated.
We denote by $\mathbb{N}$ the set of non-negative integers
and by $\mathbb{Z}$ the set of all integers.

\bigskip

\begin{definition}
\label{def1}
The \emph{$\mu$-deformed factorial function}
$\gamma _{\mu }:\mathbb{N}\rightarrow \mathbb{R}$
is defined by $\gamma _{\mu }\left( 0\right) : =1$ and
\begin{equation*}
\gamma _{\mu }\left( n\right) :=\left( n+2\mu \theta \left( n\right) \right)
\gamma _{\mu }\left( n-1\right) ,
\end{equation*}
where $n \ge 1$  and $\theta :\mathbb{N} \to \left\{ 0,1\right\} $
is the characteristic function of the odd integers.
\end{definition}

\bigskip

    This definition can be found in \cite{RO}.
In the case $\mu =0$ we obtain the known object $\gamma _{0}\left( n\right)
=n!$ (the factorial function).
Next we define the $\mu$-deformed exponential function, which also
can be found in \cite{RO}.

\bigskip

\begin{definition}
The \emph{$\mu$-deformed exponential function}
$\exp_{\mu }:\mathbb{C}\rightarrow \mathbb{C}$
is defined for $z \in \mathbb{C}$ by
\begin{equation*}
\exp_{\mu }\left( z\right) :=\sum_{n=0}^{\infty }\frac{z^{n}}{\gamma
_{\mu }\left( n\right) }.
\end{equation*}
\end{definition}

\bigskip \noindent
It is easy to see that this series converges absolutely and uniformly on compact
sets and so $\exp_{\mu }:\mathbb{C\rightarrow C}$ is
holomorphic (that is, it is an entire function).
Observe also that, since $\gamma_{0}\left( n\right) =n!$,
the undeformed exponential function $\exp_{0} $
is just the usual complex exponential function $\exp $.

\begin{theorem}
\label{thm2}
Suppose that $x\in \mathbb{R}\setminus \{0\}$.

(a) $\left\vert \exp_{\mu }(ix)\right\vert =1$
if and only if $\mu =0$.

(b) $\left\vert \exp_{\mu }(ix)\right\vert <1$
if and only if $\mu >0$.

(c) $\left\vert \exp_{\mu }(ix)\right\vert >1$
if and only if $-\frac{1}{2} <\mu <0$.
\end{theorem}

\bigskip \noindent
\textbf{Remark:} Clearly, $\exp_\mu (0) =1$ for all $\mu > -1/2$.
The implication $\Leftarrow$ of Part~(b) was proved by another
method in \cite{PS3}.

\bigskip \noindent
\textbf{Proof:}
We let $J_\nu(z)$ denote the Bessel
function of order $\nu$ with its standard domain of definition,
namely the complex plane cut along the negative real axis:
$\mathbb{C} \setminus (-\infty, 0]$.
We will use formula (3.1.2) from Rosenblum \cite{RO}:
\begin{equation}
\label{ro312}
\exp_{\mu}(-ix)=\Gamma(\mu+\tfrac{1}{2})\,2^{\mu-\frac{1}{2}}\frac{J_{\mu-\frac{1}{2}}(x)-iJ_{\mu+\frac{1}{2}}(x)}{x^{\mu-\frac{1}{2}}}.
\end{equation}
We will only need this identity for real $x > 0$.
Also, we will use the following two identities.
First, we have for all $z \in \mathbb{C} \setminus (-\infty, 0]$ that
\begin{equation}
J_{\nu-1}(z)+J_{\nu+1}(z)=\frac{2\nu}{z}J_{\nu}(z), \label{eq:abra-ste-9.1.27}
\end{equation}
which can be found as formula (9.1.27)  in \cite{AB} or
as formula (5.3.6) in \cite{LEB}.
Next, for all $z \in \mathbb{C} \setminus (-\infty, 0]$ we have that
\begin{equation}
\frac{d}{dz}\left\{ z^{-\nu}J_{\nu}(z)\right\} =-z^{-\nu}J_{\nu+1}(z),\label{eq:abra-ste-9.1.30}
\end{equation}
which comes from formula (9.1.30) in \cite{AB} or
formula (5.3.5)  in \cite{LEB}.
We take $x > 0$ in the following calculation.
(Justifications of the steps are given afterwards.)
\[
\begin{aligned}
\frac{d}{dx} & \left\{ |\exp_{\mu}(-ix)|^{2}\right\} \\
 & =\frac{d}{dx}\left\{
 \Gamma(\mu+\tfrac{1}{2})^{2}\,2^{2\mu-1}\left[\left(x^{\frac{1}{2}-\mu}
J_{\mu-\frac{1}{2}}(x)\right)^{2}+\left(x^{\frac{1}{2}-\mu}J_{\mu+\frac{1}{2}}(x)
\right)^{2}\right]\right\} \\
 &
 =\Gamma(\mu+\tfrac{1}{2})^{2}\,2^{2\mu-1}\Bigg[2\left(x^{\frac{1}{2}-\mu}
J_{\mu-\frac{1}{2}}(x)\right)\left(-x^{\frac{1}{2}-\mu}
J_{\mu+\frac{1}{2}}(x)\right)\\
 &
 \phantom{=}+2\left(x^{\frac{1}{2}-\mu}J_{\mu+\frac{1}{2}}(x)\right)
\left(-x^{\frac{1}{2}-\mu}J_{\mu+\frac{3}{2}}(x)+x^{-\mu-\frac{1}{2}}
J_{\mu+\frac{1}{2}}(x)\right)\Bigg]\\
 &
 =2^{2\mu}\Gamma(\mu+\tfrac{1}{2})^{2}x^{1-2\mu}
J_{\mu+\frac{1}{2}}(x)
\left[-J_{\mu-\frac{1}{2}}(x)-J_{\mu+\frac{3}{2}}(x)+\frac{1}{x}
J_{\mu+\frac{1}{2}}(x)\right]\\
 &
 =2^{2\mu}\Gamma(\mu+\tfrac{1}{2})^{2}x^{1-2\mu}
J_{\mu+\frac{1}{2}}(x)\left[-\frac{2(\mu+\frac{1}{2})}{x}
J_{\mu+\frac{1}{2}}(x)+\frac{1}{x}J_{\mu+\frac{1}{2}}(x)\right]\\
 &
 =(-\mu)2^{2\mu+1}\Gamma(\mu+\tfrac{1}{2})^{2}x^{-2\mu}
\left(J_{\mu+\frac{1}{2}}(x)\right)^{2}
\end{aligned}
\]
The first equality follows from equation (\ref{ro312}) and
the fact that $J_\nu(x)$ is real for $x>0$.
For the second equality we used equation
\eqref{eq:abra-ste-9.1.30} twice together with the identity
$$
 x^{\frac{1}{2}-\mu}J_{\mu+\frac{1}{2}}(x) =
   x \left( x^{-\frac{1}{2}-\mu}J_{\mu+\frac{1}{2}}(x) \right).
$$
The third and fifth equalities follow from simple algebra, while the fourth
is an application of equation \eqref{eq:abra-ste-9.1.27}.

So, for $x > 0$ the derivative of $|\exp_{\mu}(-ix)|^{2}$
has the same sign as $-\mu$ or is zero.
Since $\phi(x) := |\exp_{\mu}(-ix)|^{2} = \exp_\mu (-ix) \exp_\mu (ix)$
is an even function of $x \in \mathbb{R}$, it follows that
its derivative $\phi^\prime(x) $ is an odd function of $x \in \mathbb{R}$.
So, for $x <0$ the derivative $\phi^\prime(x) $
has the same sign as $\mu$ or is zero.
Of course, this agrees with the classical result when $\mu=0$,
namely that the derivative of
$$
|\exp_0(-ix)|^{2} = |\exp(-ix)|^{2} = 1
$$
is identically zero.
We now consider the case when $\mu \ne 0$.
Then $\phi(x) = \exp_\mu (-ix) \exp_\mu (ix)$ is
clearly real analytic (in the variable $x \in \mathbb{R}$)
and not constant.
And this implies that the critical points of $\phi(x) $
are isolated.
But $x=0$ is a critical point of $\phi(x)$, since $\phi^\prime(x)$ is odd
and continuous, implying that $\phi^\prime(0) = 0$.
And the corresponding critical value is $\phi (0) = |\exp_{\mu}(0)|^{2}= 1$.

Now the above analysis of the sign of the derivative of $|\exp_{\mu}(-ix)|^{2}$
in the intervals $(-\infty,0)$ and $(0, \infty)$ shows that the
critical value $1$ at $x=0$ is an absolute minimum if $-1/2 < \mu < 0$
while it is an absolute maximum if $\mu >0$.
And thus we have shown all three parts of the statement of the theorem.
\hfill Q.E.D.

\section{Main Result}

We are now ready to state and prove our main result.

\begin{theorem}
Let $A$ and $B$ be bounded
Borel subsets of $\mathbb{R}$ with $0\notin A^{-}$, the closure of $A$.
If $-1/2 < \mu < 0$ and $m_{\mu }(A) m_{\mu }(B)\neq 0$ then we have that
\begin{equation}
Tr\left( E^{Q_{\mu }}(A)E^{P_{\mu }}(B)\right) > m_{\mu }(A) m_{\mu }(B).
\end{equation}

In particular, the operators $Q_{\mu}$ and $P_{\mu}$
are not Accardi complementary for $-1/2 < \mu < 0$.
\end{theorem}
{\bf Proof:}
The formula (\ref{finite_trace}) of Theorem \ref{thm1} holds.
So we use the lower bound of part~(c) of Theorem \ref{thm2} to estimate
the integral in formula (\ref{finite_trace}) from below.
This gives the result.
\hfill Q.E.D.

\section{Some Identities}

\bigskip

In this section we always will take $\mu >-\frac{1}{2}$.
Recall that the $\mu$-deformed factorial function
$\gamma_\mu(n)$ has been
defined in Definition \ref{def1}.

\bigskip

\begin{definition}
The \emph{$\mu $-deformed binomial coefficient}
is defined for all $n\in \mathbb{N}$ and $k\in \mathbb{Z}$ by
\begin{equation*}
\binom{n}{k}_{\mu }:=\frac{\gamma _{\mu }(n)}{\gamma _{\mu }(n-k)\gamma
_{\mu }(k)}
\end{equation*}
if $0\leq k\leq n$ and $\binom{n}{k}_{\mu }:=0$ for other integer values of $k$.
For $n\in \mathbb{N}$ and $x,y\in \mathbb{C}$ the
\emph{$n$-th $\mu$-deformed binomial polynomial} (or
\emph{$\mu$-deformed binomial polynomial of degree} $n$) is defined by
\begin{equation*}
p_{n,\mu }\left( x,y\right) :=\sum_{k=0}^{n}\binom{n}{k}_{\mu }x^{k}y^{n-k}.
\end{equation*}
\end{definition}

\bigskip

    These definitions can be found in \cite{RO}.
In the case $\mu =0$ we obtain the known objects $\gamma _{0}\left( n\right)
=n!$ (the factorial function), $\binom{n}{k}_{0}=\binom{n}{k}$ (the binomial
coefficient), and $p_{n,0}(x,y)=\sum_{k=0}^{n}\binom{n}{k}x^{k}y^{n-k}$
(the $n$-th binomial polynomial $\left( x+y\right) ^{n}$).
Note that $p_{0,\mu }(x,y)=1$ and $p_{1,\mu }(x,y)=x+y$.
So the $\mu $-deformed binomial polynomials of degree $0$ and $1$
are the same as the
undeformed binomial polynomials of the same degree.
However, $p_{n,\mu }\left( x,y\right)$ does depend on $\mu$ for $n \ge 2$.

    Clearly we have that
$\gamma _{\mu }\left( n\right) >0$ for all $n\in \mathbb{N}$ and
thus $\binom{n}{k}_{\mu }\geq 0$ for all $n\in \mathbb{N}$ and $k\in \mathbb{Z}$.
Observe also that for all $n\in \mathbb{N}$ and
$\mu >-\frac{1}{2}$ we have that
\begin{equation*}
\binom{n}{0}_{\mu }=\binom{n}{n}_{\mu }=1
\end{equation*}
and
\begin{equation*}
\binom{n}{k}_{\mu }=\binom{n}{n-k}_{\mu }.
\end{equation*}
The Pascal Triangle property $\binom{n}{k-1}+\binom{n}{k}=\binom{n+1}{k}$
for the binomial coefficients has
the following form in the $\mu $-deformed setting.

\bigskip

\begin{theorem}
For $n\in \mathbb{N}$ and
$k\in \mathbb{Z}$ we have that
\begin{equation}
\label{2.1}
\binom{2n}{k-1}_{\mu }+\binom{2n}{k}_{\mu }=\binom{2n+1}{k}_{\mu }
\end{equation}
and
\begin{equation}
\label{2.2}
\binom{2n+1}{k-1}_{\mu }+\binom{2n+1}{k}_{\mu }=\left( 1+\frac{2\mu \theta
\left( k\right) }{n+1}\right) \binom{2n+2}{k}_{\mu }
\end{equation}
\end{theorem}

\bigskip

\textbf{Proof:}
Observe that formula (\ref{2.1}) is trivial if $k\leq 0$ or $k\geq 2n+1$.
So let us take $0<k<2n+1$.
Since $\theta \left( 2n+1-k\right) +\theta \left( k\right) =1$, we have that

\begin{eqnarray*}
&&\binom{2n}{k-1}_{\mu }+\binom{2n}{k}_{\mu } \\
&=&\frac{\gamma _{\mu }\left( 2n\right) }{\gamma _{\mu }\left( k-1\right)
\gamma _{\mu }\left( 2n-k+1\right) }+\frac{\gamma _{\mu }\left( 2n\right) }{
\gamma _{\mu }\left( k\right) \gamma _{\mu }\left( 2n-k\right) } \\
&=&\frac{k+2\mu \theta \left( k\right) }{2n+1+2\mu }\frac{\gamma _{\mu
}\left( 2n+1\right) }{\gamma _{\mu }\left( k\right) \gamma _{\mu }\left(
2n-k+1\right) } \\
&&+\frac{2n+1-k+2\mu \theta \left( 2n+1-k\right) }{2n+1+2\mu }\frac{\gamma
_{\mu }\left( 2n+1\right) }{\gamma _{\mu }\left( k\right) \gamma _{\mu
}\left( 2n-k+1\right) } \\
&=&\frac{2n+1+2\mu \left( \theta \left( 2n+1-k\right) +\theta \left(
k\right) \right) }{2n+1+2\mu }\binom{2n+1}{k}_{\mu } \\
&=&\binom{2n+1}{k}_{\mu },
\end{eqnarray*}
which proves (\ref{2.1}).
Similarly, formula (\ref{2.2}) is trivial if $k\leq 0$ or 
$k\geq 2n+2$. So let us take $0<k<2n+2$. 
Since $\theta \left( 2n+2-k\right) =\theta \left( k\right) $, we have that

\begin{eqnarray*}
&&\binom{2n+1}{k-1}_{\mu }+\binom{2n+1}{k}_{\mu } \\
&=&\frac{\gamma _{\mu }\left( 2n+1\right) }{\gamma _{\mu }\left( k-1\right)
\gamma _{\mu }\left( 2n+2-k\right) }+\frac{\gamma _{\mu }\left( 2n+1\right) 
}{\gamma _{\mu }\left( k\right) \gamma _{\mu }\left( 2n+1-k\right) } \\
&=&\frac{k+2\mu \theta \left( k\right) }{2n+2}\frac{\gamma _{\mu }\left(
2n+2\right) }{\gamma _{\mu }\left( k\right) \gamma _{\mu }\left(
2n+2-k\right) } \\
&&+\frac{2n+2-k+2\mu \theta \left( 2n+2-k\right) }{2n+2}\frac{\gamma _{\mu
}\left( 2n+2\right) }{\gamma _{\mu }\left( k\right) \gamma _{\mu }\left(
2n+2-k\right) } \\
&=&\left( 1+\frac{\mu \left( \theta \left( k\right) +\theta \left(
2n+2-k\right) \right) }{n+1}\right) \binom{2n+2}{k}_{\mu } \\
&=&\left( 1+\frac{2\mu \theta \left( k\right) }{n+1}\right) \binom{2n+2}{k}_{\mu },
\end{eqnarray*}
which proves (\ref{2.2}).
\hfill \textbf{Q.E.D.}

\bigskip

In the undeformed case we have
$\left( x+y\right) \left( x+y\right)^{n}=\left( x+y\right) ^{n+1}$.
But, when working with $\mu $-deformed
binomial polynomials $p_{n,\mu }\left( x,y\right) $ for $\mu \neq 0$, the
corresponding result is described in the following proposition.

\bigskip
\begin{theorem}
Let $x,y\in \mathbb{C}$ and $n\in \mathbb{N}$.
Then we have that

\begin{equation}
\label{2.3}
p_{1,\mu }\left( x,y\right) p_{2n,\mu }\left( x,y\right) =p_{2n+1,\mu
}\left( x,y\right)
\end{equation}
and
\begin{equation}
\label{2.4}
p_{1,\mu }\left( x,y\right) p_{2n+1,\mu }\left( x,y\right) =p_{2n+2,\mu
}\left( x,y\right) +\frac{2\mu }{n+1}\sum_{k=0}^{n}\binom{2n+2}{2k+1}_{\mu
}x^{2k+1}y^{2n+1-2k}.
\end{equation}
\end{theorem}

\bigskip

\textbf{Remark:}
Formula (\ref{2.3}) appears in Rosenblum (\cite{RO}, Corollary 4.4) where $\mu$
is assumed to be a positive parameter.

\bigskip

\textbf{Proof:} By using (\ref{2.1}) we have that

\begin{eqnarray*}
p_{1,\mu }\left( x,y\right) p_{2n,\mu }\left( x,y\right) &=&\left(
x+y\right) \sum_{k=0}^{2n}\binom{2n}{k}_{\mu }x^{k}y^{2n-k} \\
&=&\sum_{k=1}^{2n+1}\binom{2n}{k-1}_{\mu }x^{k}y^{2n-k+1}+\sum_{k=0}^{2n}
\binom{2n}{k}_{\mu }x^{k}y^{2n-k+1} \\
&=&\sum_{k=0}^{2n+1}\left( \binom{2n}{k-1}_{\mu }+\binom{2n}{k}_{\mu
}\right) x^{k}y^{2n-k+1} \\
&=&\sum_{k=0}^{2n+1}\binom{2n+1}{k}_{\mu }x^{k}y^{2n-k+1} \\
&=&p_{2n+1,\mu }\left( x,y\right) ,
\end{eqnarray*}
which proves (\ref{2.3}).
Now, by using (\ref{2.2}) we have that

\begin{eqnarray*}
&&p_{1,\mu }\left( x,y\right) p_{2n+1,\mu }\left( x,y\right) \\
&=&\left( x+y\right) \sum_{k=0}^{2n+1}\binom{2n+1}{k}_{\mu }x^{k}y^{2n+1-k}
\\
&=&\sum_{k=1}^{2n+2}\binom{2n+1}{k-1}_{\mu }x^{k}y^{2n+2-k}+\sum_{k=0}^{2n+1}
\binom{2n+1}{k}_{\mu }x^{k}y^{2n+2-k} \\
&=&\sum_{k=0}^{2n+2}\left( \binom{2n+1}{k-1}_{\mu }+\binom{2n+1}{k}_{\mu
}\right) x^{k}y^{2n+2-k} \\
&=&\sum_{k=0}^{2n+2}\left( 1+\frac{2\mu \theta \left( k\right) }{n+1}\right) 
\binom{2n+2}{k}_{\mu }x^{k}y^{2n+2-k} \\
&=&\sum_{k=0}^{2n+2}\binom{2n+2}{k}_{\mu }x^{k}y^{2n+2-k}+\frac{2\mu }{n+1}
\sum_{k=0}^{2n+2}\theta \left( k\right) \binom{2n+2}{k}_{\mu }x^{k}y^{2n+2-k}
\\
&=&p_{2n+2,\mu }\left( x,y\right) +\frac{2\mu }{n+1}\sum_{k=0}^{n}\binom{2n+2
}{2k+1}_{\mu }x^{2k+1}y^{2n+1-2k},
\end{eqnarray*}
which proves (\ref{2.4}).
\hfill \textbf{Q.E.D.}

\bigskip

\begin{theorem}
\label{lastthm}

(a) For $n\in \mathbb{N}$ we have that
\begin{equation}
\label{2.5}
p_{2n+1,\mu }\left( 1,-1\right) =0.
\end{equation}

(b) $p_{0,\mu}(1,-1) =1$.

(c) For $n\ge 1$ we have that

\begin{equation}
\label{4.6}
p_{2n,\mu}(1,-1) = \frac{2 \mu}{n} \, \sum_{k=0}^{n-1} 
\left( \begin{array}{c} 2n \\ 2k+1 \end{array} \right)_\mu
\end{equation}

(d) For $n\in \mathbb{N}$ we have that

\begin{equation}
\label{2.6}
p_{2n,\mu }\left( 1,-1\right) =\frac{2^{2n}\mu }{n+\mu }
\prod_{k=1}^{n}\frac{k+\mu }{k+2\mu }.
\end{equation}

(e) For $n\ge 1$ we have that

\begin{equation}
\label{4.8}
p_{4n,\mu} (1,-1) =
\mu \, \frac{2^{2n} \prod_{k=n+1}^{2n-1} (\mu + k ) }{ \prod_{k=1}^{n} ( \mu + k - 1/2)  }
\end{equation}

(f) For $n\ge 1$ we have that

\begin{equation}
\label{4.9}
p_{4n-2,\mu} (1,-1) = \mu \, \frac{2^{2n-1} \prod_{k=n+1}^{2n-1} (\mu + k - 1 ) }{ \prod_{k=1}^{n} ( \mu + k - 1/2)  } 
\end{equation}

\end{theorem}

\vskip .2cm \noindent
\textbf{Remark:} Formulas (\ref{4.6}), (\ref{4.8}) and (\ref{4.9}) were
conjectured in \cite{PS3}.
Note that (\ref{2.6}) is new and that it turns out, as we will show,
to be a compact way of writing both (\ref{4.8}) and (\ref{4.9}).

\vskip .2cm \noindent
\textbf{Proof:}
(a) Though (\ref{2.5}) is a direct consequence of (\ref{2.3})
with $x=1$ and $y=-1$ (since
$p_{2n+1,\mu }\left( 1,-1\right) =p_{1,\mu }\left( 1,-1\right) p_{2n,\mu
}\left( 1,-1\right) =\left( 1-1\right) p_{2n,\mu }\left( 1,-1\right) =0$),
we would like to mention that one can prove (\ref{2.5}) proceeding directly from
the definition and using the symmetry property
$\binom{n}{k}_{\mu }=\binom{n}{n-k}_{\mu }~$ mentioned above:

\begin{eqnarray*}
p_{2n+1,\mu }\left( 1,-1\right) &=&\sum_{k=0}^{2n+1}\binom{2n+1}{k}_{\mu
}\left( -1\right) ^{k} \\
&=&\sum_{k=0}^{n}\binom{2n+1}{k}_{\mu }\left( -1\right)
^{k}+\sum_{k=n+1}^{2n+1}\binom{2n+1}{k}_{\mu }\left( -1\right) ^{k} \\
&=&\sum_{k=0}^{n}\binom{2n+1}{k}_{\mu }\left( -1\right) ^{k}+\sum_{j=n}^{0}
\binom{2n+1}{2n+1-j}_{\mu }\left( -1\right) ^{2n+1-j} \\
&=&\sum_{k=0}^{n}\binom{2n+1}{k}_{\mu }\left( -1\right) ^{k}-\sum_{j=0}^{n}
\binom{2n+1}{j}_{\mu }\left( -1\right) ^{j} \\
&=&0.
\end{eqnarray*}

(b) This is immediate.

(c)
First we observe that using (\ref{2.4}) with $x=1$, $y=-1$ and
$n$ replaced by $n-1$, we obtain

\begin{equation*}
p_{2n,\mu }\left( 1,-1\right) +\frac{2\mu }{n}\sum_{k=0}^{n-1}\binom{2n}{2k+1
}_{\mu }(-1)=p_{1,\mu }\left( -1,1\right) p_{2n-1,\mu }\left( 1,-1\right) =0,
\end{equation*}
and therefore

\begin{equation}
\label{2.7}
p_{2n,\mu }\left( 1,-1\right) =\frac{2\mu }{n}\sum_{k=0}^{n-1}
\binom{2n}{2k+1}_{\mu }
\end{equation}
for $n \ge 1$.
This proves Part (c).

(d)
The case $n=0$ as well as the case $\mu=0$ are each trivial.
(In the case when both $n=0$ and $\mu=0$ we use the convention that
$\mu/(n+\mu) = 1 $.
We also use the standard convention that a product over an empty index set is $1$.)
So hereafter we take $n \ge 1$ and $\mu \ne 0$.

Using the previous formula we obtain for $n \ge 1$ that
\begin{eqnarray*}
p_{2n,\mu }\left( 1,-1\right) &=&\frac{2\mu }{n}\sum_{k=0}^{n-1}\binom{2n}{
2k+1}_{\mu } \\
&=&\frac{\mu }{n}\left( \sum_{k=0}^{2n}\binom{2n}{k}_{\mu }-\sum_{k=0}^{2n}
\binom{2n}{k}_{\mu }\left( -1\right) ^{k}\right) \\
&=&\frac{\mu }{n}\big( p_{2n,\mu }\left( 1,1\right) -p_{2n,\mu }\left(
1,-1\right) \big) ,
\end{eqnarray*}
and thus
\begin{equation}
\label{2.8}
p_{2n,\mu }\left( 1,-1\right) =\frac{\mu }{n+\mu }p_{2n,\mu }\left(
1,1\right) .
\end{equation}
From (\ref{2.3}) with $x=y=1$ we obtain
\begin{equation}
\label{2.9}
p_{2n+1,\mu }\left( 1,1\right) = p_{1,\mu }\left( 1,1\right) p_{2n,\mu
}\left( 1,1\right) =(1+1) p_{2n,\mu }(1,1) = 2p_{2n,\mu }\left( 1,1\right).
\end{equation}
Similarly from (\ref{2.4}) with $x=y=1$ we get
\begin{equation*}
p_{1,\mu }\left( 1,1\right) p_{2n-1,\mu }\left( 1,1\right) =p_{2n,\mu
}\left( 1,1\right) +\frac{2\mu }{n}\sum_{k=0}^{n-1}\binom{2n}{2k+1}_{\mu },
\end{equation*}
which by using $p_{1,\mu }\left( 1,1\right) =2$ and (\ref{2.7}) becomes
\begin{eqnarray*}
2 p_{2n-1,\mu }\left( 1,1\right) =p_{2n,\mu }\left( 1,1\right) +p_{2n,\mu
}\left( 1,-1\right)
\end{eqnarray*}
This last expression together with (\ref{2.8}) gives us
\begin{eqnarray*}
2p_{2n-1,\mu }\left( 1,1\right) &=&p_{2n,\mu }\left( 1,1\right) +p_{2n,\mu
}\left( 1,-1\right) \\
&=&p_{2n,\mu }\left( 1,1\right) +\frac{\mu }{n+\mu }p_{2n,\mu }\left(
1,1\right) \\
&=&\frac{n+2\mu }{n+\mu }p_{2n,\mu }\left( 1,1\right) .
\end{eqnarray*}
So we have
\begin{equation}
\label{2.10}
p_{2n,\mu }\left( 1,1\right) =\frac{2\left( n+\mu \right) }{n+2\mu }
p_{2n-1,\mu }\left( 1,1\right) .
\end{equation}
We claim that for $n\in \mathbb{N}$ we have that
\begin{equation}
\label{2.11}
p_{2n,\mu }\left( 1,1\right) =2^{2n}\prod_{k=1}^{n}\frac{k+\mu }{
k+2\mu }
\end{equation}
This is trivial for $n=0$, while for $n=1$ we have
\begin{eqnarray*}
p_{2,\mu }\left( 1,1\right) &=&\sum_{k=0}^{2}\binom{2}{k}_{\mu }=2+\binom{2}{
1}_{\mu }=2+\frac{2}{1+2\mu } \\
&=&4\frac{1+\mu }{1+2\mu }=2^{2(1)}\prod_{k=1}^{1}\frac{k+\mu }{
k+2\mu }.
\end{eqnarray*}
Arguing by induction, we now assume that
(\ref{2.11}) is valid for a given $n\in \mathbb{N}$.
Then by also using (\ref{2.9}) and (\ref{2.10}) we have
\begin{eqnarray*}
p_{2n+2,\mu }\left( 1,1\right) &=&\frac{2\left( n+1+\mu \right) }{n+1+2\mu }
p_{2n+1,\mu }\left( 1,1\right) \\
&=&\frac{2\left( n+1+\mu \right) }{n+1+2\mu }2p_{2n,\mu }\left( 1,1\right) \\
&=&\frac{2^{2}\left( n+1+\mu \right) }{n+1+2\mu }2^{2n}
\prod_{k=1}^{n}\frac{k+\mu }{k+2\mu } \\
&=&2^{2n+2}\prod_{k=1}^{n+1}\frac{k+\mu }{k+2\mu },
\end{eqnarray*}
which proves (\ref{2.11}) for $n+1$ and so proves our claim.
Finally, from (\ref{2.8}) and (\ref{2.11}) we have that
\begin{eqnarray*}
p_{2n,\mu }\left( 1,-1\right) &=&\frac{\mu }{n+\mu }p_{2n,\mu }\left(
1,1\right) \\
&=&\frac{2^{2n}\mu }{n+\mu }\prod_{k=1}^{n}\frac{k+\mu }{k+2\mu },
\end{eqnarray*}
which proves (\ref{2.6}) and so concludes the proof of Part (d).

(e) Using (\ref{2.6}) for $2n$ in place of $n$ we have that
\begin{eqnarray*}
&&p_{4n}(1,-1) = \frac{2^{4n} \mu}{2n+\mu} \prod_{k=1}^{2n} \frac{k+\mu}{k+2\mu} \\
&=& \frac{2^{2n}\mu}{2n+\mu} \prod_{k=1}^{2n} \frac{2k+2\mu}{k+2\mu}\\
&=& \frac{2^{2n}\mu}{2n+\mu} \cdot
\frac{(2+2\mu)(4+2\mu) \cdots (4n-2+2\mu)(4n+2\mu)}
{(1+2\mu)(2+2\mu)\cdots (2n-1+2\mu)(2n+2\mu)}\\
&=& \frac{2^{2n}\mu}{2n+\mu} \cdot
\frac{(2n+2+2\mu)(2n+4+2\mu) \cdots (4n-2+2\mu)(4n+2\mu)}
{(1+2\mu)(3+2\mu) \cdots (2n-3+2\mu)(2n-1+2\mu)} \\
&=& \frac{2^{2n}\mu}{2n+\mu} \cdot \frac{2^n}{2^n} \cdot
\frac{(\mu +n+1)(\mu +n+2) \cdots (\mu +2n-1)(\mu+2n)}
{(\mu +1/2)(\mu +3/2) \cdots (\mu +n-3/2)(\mu +n -1/2)} \\
&=& 2^{2n} \mu \cdot \frac{\prod_{k=n+1}^{2n-1} (\mu+k) }
{\prod_{k=1}^n (\mu +k -1/2)}
\end{eqnarray*}
and this shows (\ref{4.8}).

(f) Using (\ref{2.6}) for $2n-1$ in place of $n$ we have that
\begin{eqnarray*}
&&p_{4n-2}(1,-1) = \frac{2^{4n-2} \mu}{2n-1+\mu} \prod_{k=1}^{2n-1}
\frac{k+\mu}{k+2\mu} \\
&=& \frac{2^{2n-1}\mu}{2n-1+\mu} \prod_{k=1}^{2n-1} \frac{2k+2\mu}{k+2\mu}\\
&=& \frac{2^{2n-1}\mu}{2n-1+\mu} \cdot
\frac{(2+2\mu)(4+2\mu) \cdots (4n-4+2\mu)(4n-2+2\mu)}
{(1+2\mu)(2+2\mu)\cdots (2n-2+2\mu)(2n-1+2\mu)}\\
&=& \frac{2^{2n-1}\mu}{2n-1+\mu} \cdot
\frac{(2n+2\mu)(2n+2+2\mu) \cdots (4n-4+2\mu)(4n-2+2\mu)}
{(1+2\mu)(3+2\mu) \cdots (2n-3+2\mu)(2n-1+2\mu)} \\
&=& \frac{2^{2n-1}\mu}{2n-1+\mu} \cdot \frac{2^n}{2^n} \cdot
\frac{(\mu +n)(\mu +n+1) \cdots (\mu +2n-2)(\mu+2n-1)}
{(\mu +1/2)(\mu +3/2) \cdots (\mu +n-3/2)(\mu +n -1/2)} \\
&=& 2^{2n-1} \mu \cdot \frac{\prod_{k=n+1}^{2n-1} (\mu+k-1) }
{\prod_{k=1}^n (\mu +k -1/2)}
\end{eqnarray*}
and this shows (\ref{4.9}).

\hfill \textbf{Q.E.D.}

\section{Acknowledgments}
This article was written while two of the authors (L.A.E.C. and S.B.S.)
were on an academic visit at the University of Virginia.
They would like to thank everyone who made this possible, but most especially
Lawrence Thomas who served as their host.

\end{document}